\documentclass[a4paper,11pt]{article}

\pdfoutput=0 

\usepackage{jheppub} 

\usepackage[T1]{fontenc} 

\def\d{{\rm d}}
\def\eps{\varepsilon}

\def\beq{\begin{equation}}
\def\eeq{\end{equation}}
\def\bea{\begin{eqnarray}}
\def\eea{\end{eqnarray}}

\title{\boldmath How robust is a thermal photon interpretation of
the ALICE low-$p_T$ data?}

\author[a]{M.\ Klasen,}
\author[b,c]{C.\ Klein-B\"osing,}
\author[a]{F.\ K\"onig}
\author[b]{and J.\ P.\ Wessels}

\affiliation[a]{Institut f\"ur Theoretische Physik, Westf\"alische
 Wilhelms-Universit\"at M\"unster, \\ Wilhelm-Klemm-Stra\ss{}e 9,
 D-48149 M\"unster, Germany}
\affiliation[b]{Institut f\"ur Kernphysik, Westf\"alische
 Wilhelms-Universit\"at M\"unster, \\ Wilhelm-Klemm-Stra\ss{}e 9,
 D-48149 M\"unster, Germany}
\affiliation[c]{
ExtreMe Matter Institute, GSI,
Planckstra\ss{}e 1, D-64291 Darmstadt,
Germany}

\emailAdd{michael.klasen@uni-muenster.de}
\emailAdd{c.klein-boesing@uni-muenster.de}
\emailAdd{f.koenig@uni-muenster.de}
\emailAdd{j.wessels@uni-muenster.de}

\preprint{MS-TP-13-18}

\abstract{We present a rigorous theoretical analysis of the ALICE measurement
 of low-$p_T$ direct-photon production in central lead-lead collisions at the LHC
 with a centre-of-mass energy of $\sqrt{s_{NN}}=2.76$ TeV.  Using NLO QCD, we compute
 the relative contributions to prompt-photon production from
 different initial and final states and the theoretical uncertainties coming
 from independent variations of the renormalisation and factorisation scales,
 the nuclear parton densities and the fragmentation functions. Based on
 different fits to the unsubtracted and prompt-photon subtracted ALICE data,
 we consistently find $T=304$ $\pm$ 58 MeV and $309\pm64$ MeV for the effective
 temperature of the quark-gluon plasma (or hot medium) at $p_T\in[0.8;2.2]$ GeV and
 $p_T\in[1.5;3.5]$ GeV as well as a power-law ($p_T^{-4}$) behavior for $p_T>4$
 GeV as predicted by QCD hard scattering.}

\begin{document} 
\maketitle
\flushbottom

\section{Introduction}

One of the goals of the physics program at the CERN Large Hadron Collider (LHC)
is the study of deconfined, strongly interacting matter, which existed in the
early universe and which can today be re-created and investigated in heavy-ion
collisions. An important probe for this so-called quark-gluon plasma (QGP) are
photons emitted from the deconfined partons before thermalisation, in the
thermal bath, during expansion and cooling of the QGP, and finally from the thermal
hadron gas. The transverse momentum distribution of these photons can be used to
estimate the temperature of the system, although the exact interpretation is
complicated by these different phases, the radial expansion, more generally
the temporal evolution of phase space, and co-existing states of matter. Experimentally,
an effective temperature is usually extracted from an exponential 
fit to the excess of direct-photon production at low transverse momentum
($p_T$) above the expectation from vacuum production.

In Ref.\ \cite{Wilde:2012wc}, the first observation of a low-$p_T$ direct-photon
signal at the LHC has been reported by the ALICE collaboration.\footnote{In nuclear
collision experiments, photons originating from meson decays are usually
distinguished from {\em direct} photons. The latter are divided into {\em thermal} and
non-thermal {\em prompt} (plus medium-induced)
photons, and these again into photons produced {\em directly}
in the hard collision and those coming from quark and gluon {\em fragmentation}.
The double use of the word {\em direct} can sometimes lead to confusion.}
There, an inverse slope parameter of $T_{\rm LHC}=304\pm51$ MeV has been
extracted from an exponential
fit to the photon spectrum in central (0-40\%) lead-lead
collisions at $\sqrt{s_{NN}}=2.76$ TeV and low transverse momenta of
$p_T\in[0.8;2.2]$ GeV. The inverse slope parameter of this
measurement is significantly higher than the one obtained previously
by the PHENIX collaboration in 0-20\% central gold-gold collisions with
$\sqrt{s_{NN}}=200$ GeV at RHIC, which resulted in $T_{\rm RHIC}=221\pm19\pm19$ MeV
\cite{Adare:2008ab}. The latter was higher than
the transition temperature to the QGP of about $T_{\rm crit}=170$ MeV, but
1.5 to 3 times smaller
than the initial temperature $T_0$ of the dense matter due to the
space-time evolution following its initial formation;
in hydrodynamical models, which describe the data at $\sqrt{s_{NN}}$ =
200 GeV, $T_0$ ranges from 600 to 300 MeV depending on the
formation time, assumed to lie between $\tau_0 =  0.15$ and
0.6 fm$/c$ \cite{d'Enterria:2005vz}.

The two general-purpose LHC experiments ATLAS and CMS have also recently
measured prompt-photon production in lead-lead collisions at $\sqrt{s_{NN}}=2.76$ TeV
with $p_T > 40$ and 20 GeV, respectively \cite{ATLAS:2012zla,Chatrchyan:2012vq}.
In this transverse momentum range, photons (like electroweak $Z$ and $W$ bosons
\cite{Steinberg:2013laa,GranierdeCassagnac:2013maa})
are expected to be dominantly produced in hard partonic collisions
and to be unaffected by the strongly interacting medium, so that they
can be described with next-to-leading order (NLO) QCD programs such as
JETPHOX \cite{Aurenche:2006vj}.

However, photons from
electromagnetic decays of neutral mesons (mostly $\pi^0,\eta\to\gamma\gamma$)
must first be removed using meson data \cite{Wilde:2012wc}, Monte Carlo simulations
\cite{Adare:2008ab} and/or by applying an isolation criterion, which e.g.\
limits the energy of all other particles
in a cone of size $R_{\rm iso}=\sqrt{\Delta\eta^2+\Delta\phi^2}<0.3$
or 0.4 around the photon to $E_T(R_{\rm iso})<6$ or 5 GeV 
\cite{ATLAS:2012zla,Chatrchyan:2012vq}. 
The remaining ``prompt'' photon sample then still contains both direct
photons, produced at tree-level through the processes $q\bar{q}\to\gamma g$
and $qg\to\gamma q$, and photons emitted in the collinear fragmentation of
quarks and gluons produced in partonic scattering processes.
Isolation criteria also reduce the contribution of fragmentation photons,
since these are generally produced inside a high-$p_T$ jet of hadrons.
The separation of these direct and fragmentation photons becomes ambiguous, i.e.\
scheme- and scale-dependent, at NLO of perturbative QCD, but their
interplay and the factorisation of the associated infrared singularities
have been studied intensively theoretically \cite{Klasen:2002xb}. As a
result, NLO calculations describe very well prompt-photon production in
$pp$ collisions at the LHC with $\sqrt{s}=7$ TeV down to $p_T \geq 21$
GeV \cite{Aad:2011tw,ATLAS:2013ema,Khachatryan:2010fm,Chatrchyan:2011ue}.

At smaller transverse momenta, a transition from hard to thermal photon production
is expected to occur in central heavy-ion collisions, in contrast to peripheral
heavy-ion and $pp$ collisions. For this reason,
the transverse momentum distribution of inclusive photons
has been measured by the ALICE collaboration
in lead-lead collisions at $\sqrt{s_{NN}}=2.76$ TeV from 16 GeV down
to 0.4 GeV. In addition, the same has been done in $pp$ collisions at $\sqrt{s}=7$ TeV down
to 0.3 GeV. In both cases, contributions from decay photons were subtracted
using neutral pion data and
transverse mass ($m_T$) scaling \cite{Khandai:2011cf}.
The data for the 0-40\% most central heavy-ion collisions showed a clear
deviation from the perturbative QCD prediction below $p_T$-values of 4 GeV
which was not observed for $pp$ and peripheral (40-80\%) heavy-ion collisions
\cite{Wilde:2012wc}.
Similar observations had been made in a preceding analysis of the PHENIX
collaboration in minimum bias, 0-20\% and 20-40\% central gold-gold
collisions at RHIC with $\sqrt{s_{NN}}=200$ GeV \cite{Adare:2008ab}.

In this paper we will scrutinise the thermal photon interpretation
of the ALICE low-$p_T$ data with respect to theoretical uncertainties in
the perturbative QCD prediction.
Not only do these uncertainties arise from variations of the renormalisation and
factorisation scales in the incoming and outgoing hadrons and photons,
which have so far only been varied simultaneously by a factor of two
about the photon $p_T$, but they are also caused by uncertainties in the factorisation
scheme \cite{Klasen:1996yk}, the nuclear parton density functions (nPDFs)
and the photon fragmentation functions (FFs).

Nuclear PDFs are subject to intrinsic uncertainties of the proton PDFs, in particular
of the gluon \cite{Stump:2003yu}, and cold nuclear modification effects,
such as nuclear shadowing \cite{Hammon:1998gq}. Both are expected to be
significantly larger at the LHC with its small values of partonic momentum
fractions $x\simeq 2p_T/\sqrt{s_{\rm NN}}\simeq10^{-3}$ than at RHIC and at low scales of
$\mu_f\simeq p_T\simeq 2$ GeV rather than at large scales.
Extra $p_T$-broadening (the Cronin effect)
is expected to become small at high (RHIC and LHC) centre-of-mass energies and small $x$
\cite{Dumitru:2001jx}. The latter has been confirmed experimentally at RHIC by the
measurement of direct photons in $d$+Au reactions
\cite{Adare:2012vn} and at the LHC via the comparison of charged hadron
production in proton-proton and proton-nucleus collisions
\cite{ALICE:2012mj}.
In general, when going from proton-proton to nucleus-nucleus collisions,
one must also account for isospin suppression of
prompt-photon production due the presence of neutrons (126) and
not only protons (82) in the lead nucleus.
This is, however, mostly relevant in the valence-quark region at large $x$.
Saturation effects appear to set in only at very small
values of $x$. The influence of these effects has been analysed in several
global fits of nPDFs \cite{Eskola:2009uj,Hirai:2007sx,deFlorian:2011fp,Schienbein:2009kk}.

The photon fragmentation functions are also subject to considerable
uncertainties, in particular the one for gluons. In the absence of
sufficiently detailed experimental data from $e^+e^-$ collisions,
they are typically modeled on vector meson fragmentation
\cite{Bourhis:1997yu}. As an alternative, it has been proposed to
use slightly virtual photons decaying into low-mass lepton pairs,
as these do not have fragmentation contributions and need not be isolated
experimentally \cite{Berger:1998ev,Berger:1999es,Brandt:2013hoa}.

The remainder of this paper is organised as follows: In Sec.\
\ref{sec:2}, we will study the relative importance of the different
partonic subprocesses contributing to prompt photon production
in lead-lead collisions. Sections \ref{sec:3} and \ref{sec:4} give
a brief review of the current uncertainty on the nuclear
PDFs and photon FFs. Our main results are presented in Sec.\
\ref{sec:5}, where we compare our NLO QCD predictions to the
ALICE data and estimate the total theoretical uncertainty coming
from the various sources defined above. Our conclusions are given
in Sec.\ \ref{sec:6}. For future use, we present in App.\ \ref{app:a}
our NLO QCD calculations for the invariant prompt-photon yields
in tabular form. 

\section{Partonic subprocesses}
\label{sec:2}

As stated in the introduction, the goal of this paper is to test
how robust a thermal photon interpretation of the low-$p_T$
ALICE data is. We will explore to what extent the observed
excess of photons with transverse momenta below 4 GeV can be
explained by the production of hard photons in the initial
partonic collision alone, taking into account all theoretical
uncertainties.

The production of these photons is calculated perturbatively in NLO QCD
using JETPHOX 1.2 \cite{Aurenche:2006vj}. As is well known, the truncation of the
perturbative series at this order leads to an artificial dependence on
the unphysical renormalisation scale $\mu$ and the initial- and
final-state factorisation scales $\mu_f$ and $\mu_D$. These
have so far been varied only simultaneously, and we will later
vary them also individually by a factor of two around the
central scale $p_T$.
In addition, uncertainties in the determination of the PDFs
must also be taken into account, and in the case of heavy-ion
collisions not only for the free-proton PDFs, but also for their
nuclear modification, which is commonly parametrised as a modification
factor $R_{i/A}(x,\mu_f)$ of the density for parton $i$.
Furthermore, at low transverse momenta and in the absence of
a photon isolation criterion, fragmentation processes are expected
to dominate, thus introducing an additional uncertainty from
the photon fragmentation function.

Before we address these sources of uncertainty in the
following chapters, it is illustrative to learn about the
relative contributions of the various partonic subprocesses
leading to prompt-photon production in lead-lead collisions.
These depend in particular on the momentum fractions $x$ and
$z$ of the partons in the PDFs $f(x,\mu_f)$ and FFs $D(z,\mu_D)$,
but also on the factorisation scales $\mu_f$ and $\mu_D$.
Since we are interested in the low-$p_T$ region, we set these
scales here to 2 GeV, which is already in the perturbative
regime, but only slightly larger than the starting scale
$Q_0\sim1.3$ GeV \cite{Eskola:2009uj,Hirai:2007sx,deFlorian:2011fp,Schienbein:2009kk}.
The relevant momentum fractions in the
initial state can be computed from the photon's transverse
momentum by $x_T=2p_T/\sqrt{s_{NN}}$ (the true $x$ at central
rapidity),
leading to typical values of $x_T\sim10^{-3}$. Our simulations
show that the momentum fractions in the final-state fragmentation
range from $z\sim0.01$ ($0.1$) to 1 at $p_T=2$ (16) GeV.
In this section, we use the best fit of EPS09 \cite{Eskola:2009uj}
for the nPDFs and the BFG set II \cite{Bourhis:1997yu} for the FFs
as our baseline.

%
\begin{figure}
 \centering
 \epsfig{file=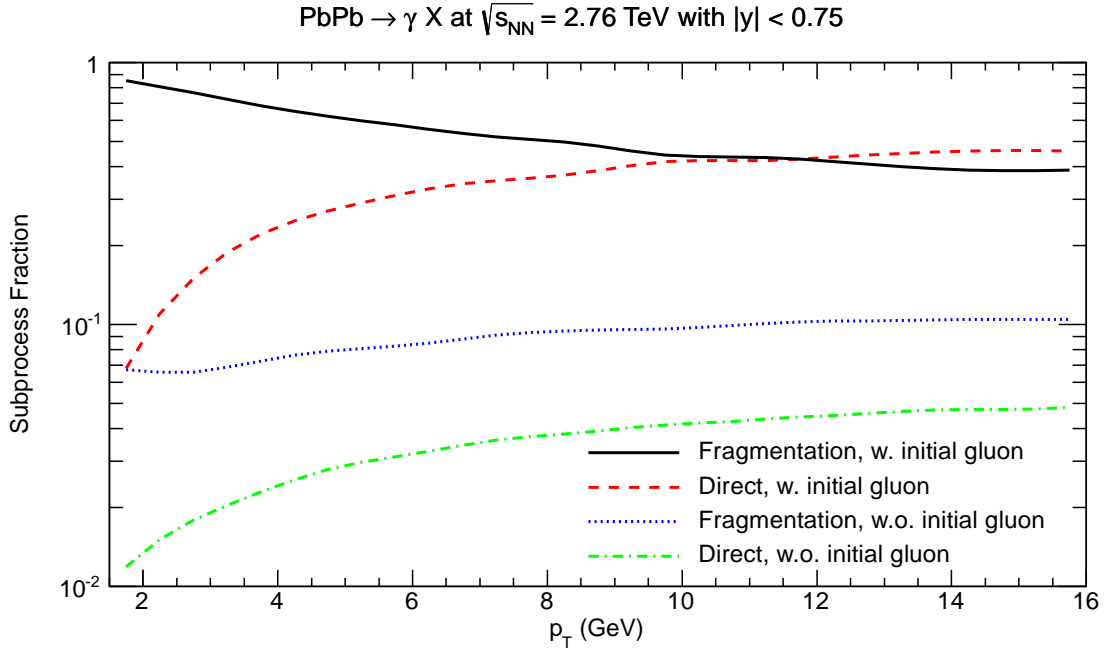,width=\columnwidth}
 \caption{\label{fig:1}Relative contributions to the total
 NLO prompt-photon cross section in lead-lead collisions with
 a center-of-mass energy of
 $\sqrt{s_{NN}}=2.76$ TeV. We show individually the fragmentation
 (full) and direct (dashed) subprocesses with initial gluons as well as
 those with initial quarks and antiquarks only (dotted and dot-dashed).}
\end{figure}
%
In Fig.\ \ref{fig:1}, we plot the 
relative contributions to the total NLO prompt-photon
cross section in lead-lead collisions with $\sqrt{s_{NN}}=2.76$ TeV.
Individually shown are the fragmentation (full) and direct (dashed)
subprocesses with initial gluons as well as those with initial
quarks and antiquarks only (dotted and dot-dashed). As expected,
fragmentation contributions dominate at low transverse momenta (up to
12 GeV),
i.e.\ over almost the entire $p_T$-range measured by ALICE.
Only at larger transverse momenta the direct QCD ``Compton'' process
$qg\to \gamma g$ takes over. In the initial state, the gluon contribution
dominates in fact throughout, while pure quark and antiquark initial states
make up for at most ten percent. We therefore expect the uncertainty on the
gluon in bound nucleons at small $x$ to play an important role.

\begin{figure}
 \centering
 \epsfig{file=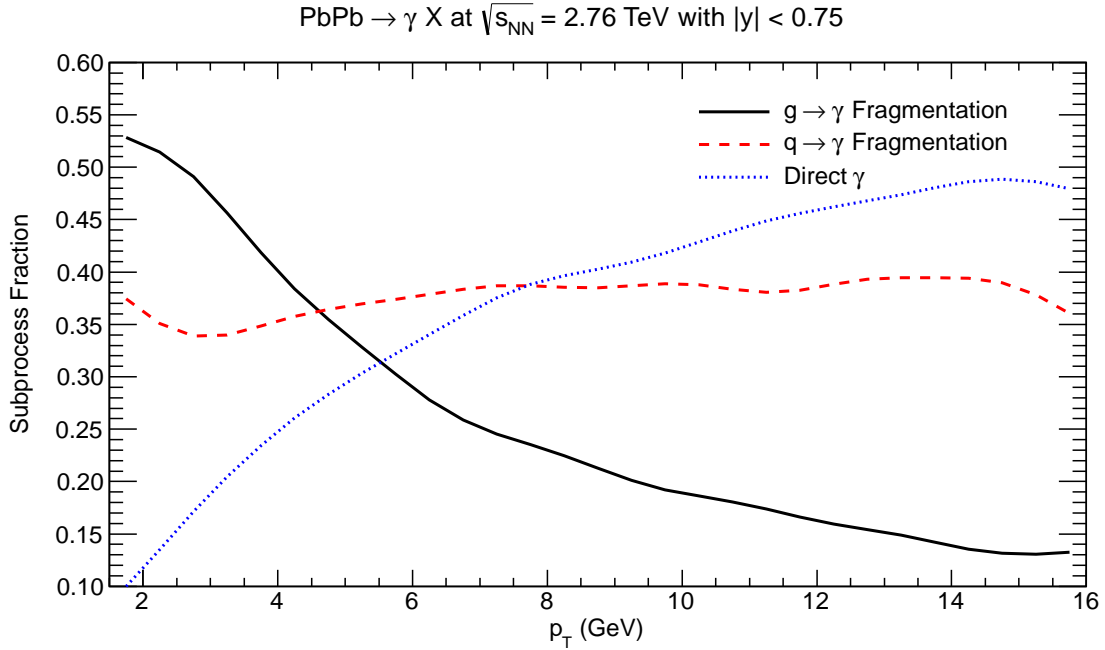,width=\columnwidth}
 \caption{\label{fig:2}Relative final-state contributions to the total
 NLO prompt-photon cross section in lead-lead collisions with
 $\sqrt{s_{NN}}=2.76$ TeV. We show individually the gluon (full) and
 quark (dashed) fragmentation subprocesses as well as the direct
 subprocesses (dotted).}
\end{figure}

Looking at the final state, we plot in Fig.\ \ref{fig:2}
%
%
the subprocesses with gluon (full) and quark (dashed) fragmentation as well
as the direct subprocesses (dotted). As can be seen, the very low-$p_T$
region (below 4 GeV) is dominated by gluon fragmentation into photons,
which is very poorly constrained experimentally.
Quark fragmentation gives an almost constant contribution of about 40\%
over the entire $p_T$-range. Direct photons become larger than quark
fragmentation at transverse momenta of 8 GeV and than all fragmentation
processes above 12 GeV (see above).

\section{Nuclear parton density uncertainties}
\label{sec:3}

In this paper, we use the EPS09 nPDFs as our baseline \cite{Eskola:2009uj}.
There, the bound-proton PDFs are defined in terms of nuclear
modifications $R_{i/A}(x,\mu_f)$ applied to CTEQ6.1M \cite{Stump:2003yu}
free-proton PDFs $f_{i/p}(x,\mu_f)$,
\beq
 f_{i/A}(x,\mu_f) \equiv R_{i/A}(x,\mu_f) \cdot f_{i/p}(x,\mu_f).
 \label{eq:1}
\eeq
The bound-neutron PDFs are obtained by assuming isospin symmetry.
E.g.\ the total up-quark ($u$) distribution per nucleon in a nucleus $A$ with $Z$
protons is
\beq
 f_{u/A}(x,\mu_f) = {Z\over A} [R_{u/A}^v f^v_{u/p} + R^s_{u/A} f^s_{u/p}] +
            {A-Z\over A}  [R_{d/A}^v f^v_{d/p} + R^s_{d/A} f^s_{d/p}],
 \label{eq:2}
\eeq
where $d$ corresponds to the down-quark and
the superscripts $v$ and $s$ refer to valence and sea quark contributions,
respectively. The parametrisation of the nuclear modifications $R_{i/A}(x,\mu_f)$
is performed at the charm quark mass ($m_c$) threshold imposing the
momentum and baryon number sum rules
\beq
 \sum_{i=q,\overline{q},g} \int_0^1 dx \, xf_{i/A}(x,m_c) = 1, \quad \int_0^1 dx \left[ f^v_{u/A}(x,m_c) + f^v_{d/A}(x,m_c) \right] = 3, \label{eq:sumrules}
\eeq
for each nucleus $A$ separately. At higher scales, 
the nPDFs are obtained by solving the DGLAP evolution equations. This
approach results in an excellent fit to the different types of nuclear hard-process
data \cite{Eskola:2009uj}, suggesting that factorisation works well in the energy
range studied and that
the extracted nPDFs are universal in the region $x\geq0.005$, $\mu_f\geq 1.3$ GeV.

The theoretical uncertainties of the free-proton PDFs are obtained using the 40 error
sets of the CTEQ6.1M parametrisation \cite{Stump:2003yu}.
An additional 30 error sets are assigned pairwise to the uncorrelated eigendirections
of the 15 parameters fitted to the nuclear collision data sets. A total uncertainty band
at 90\% confidence level is then calculated from the 71 sets defined by
fixing either $R_{i/A}(x,\mu_f)$  to the best fit value and varying the
free-proton PDFs or fixing the latter to its best fit value and varying the former.
These variations then contribute pairwise to the size of the upper and lower errors via
\bea
 \delta^+f&=&\sqrt{\sum_{i}[\max(f_i^{(+)}-f_0,f_i^{(-)}-f_0,0)]^2},\\
 \delta^-f&=&\sqrt{\sum_{i}[\max(f_0-f_i^{(+)},f_0-f_i^{(-)},0)]^2}.
\eea
As the authors acknowledge, this factorised approach represents a
simplification, violating, e.g., in some cases momentum conservation,
so that strictly speaking the free- and bound-proton PDF uncertainty
analyses should not be separated \cite{Eskola:2009uj}.

Note that the minimal value of $x=0.005$ constrained by experimental data in
the EPS09 fit lies above the value of $x=0.001$ expected to be relevant for the
low-$p_T$ ALICE data. The nPDFs therefore rely in this region on an extrapolation
from higher $x$-values. Experimental data down to $x=5\cdot10^{-5}$ exist from the
BRAHMS collaboration \cite{Arsene:2004ux}, who have measured charged-hadron
production at forward rapidity $\eta=2.2$ and 3.2 in $pp$ and $d$+Au collisions
at $\sqrt{s_{NN}}=200$ GeV. However, they have not been included in any of the nPDFs
and would in fact lead to a poor description by the EPS09 fits, in particular for
transverse momenta below 2 GeV at $\eta=2.2$ and below 4 GeV at $\eta=3.2$
(see Fig.\ 14 in \cite{Eskola:2009uj}).

In order to estimate the bias from different underlying
free-proton PDFs, parametrisations of the nuclear modification,
and fitted nuclear data sets, we also study the best fits of the
HKN07 \cite{Hirai:2007sx}, DSSZ \cite{deFlorian:2011fp} and nCTEQ
\cite{Schienbein:2009kk} collaborations. In particular, the HKN07
nPDFs are based on the MRST1998 free-proton set \cite{Martin:1998sq}
and those of DSSZ on the more recent MSTW2008 set \cite{Martin:2009iq}.
The nCTEQ parametrisation is (so far) the only one that does not rely on
a factorisation into a nuclear modification factor and free-proton PDFs.
Instead it introduces an explicit $A$-dependence in the coefficients
of the $x$-dependent functional form of the PDFs at the starting scale.
So only the technical framework of the CTEQ6M analysis is used here
\cite{Pumplin:2002vw}.

%
\begin{figure}
 \centering
 \epsfig{file=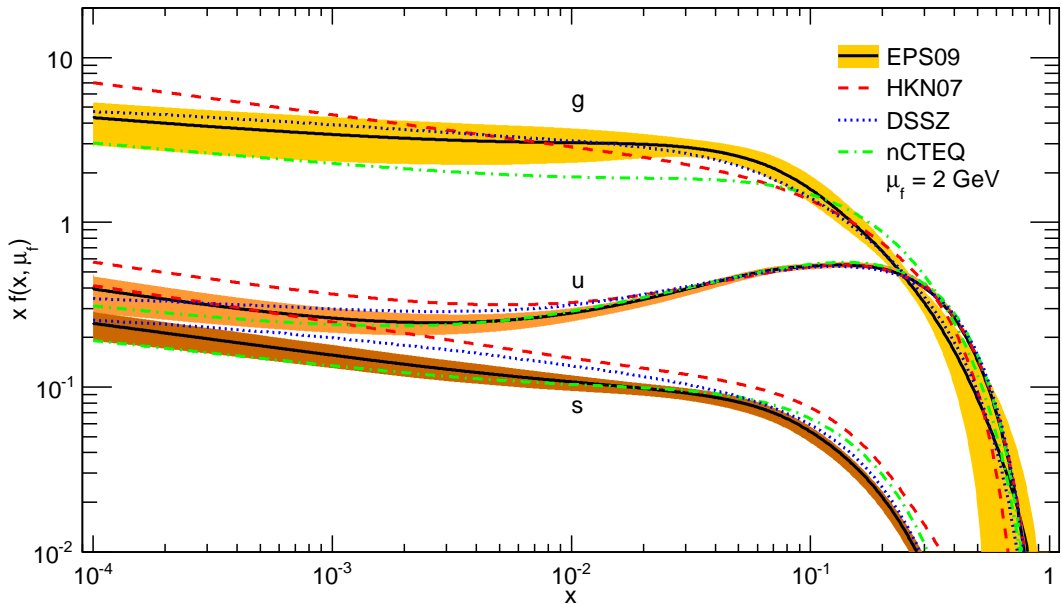,width=\columnwidth}
 \caption{\label{fig:3}Nuclear parton distribution functions (nPDFs) of gluons,
 up ($\sim$ down) and strange quarks in lead ions from the best fits of 
 different collaborations \cite{Eskola:2009uj,Hirai:2007sx,deFlorian:2011fp,Schienbein:2009kk},
 at the factorisation scale $\mu_f = 2$ GeV. The uncertainty band is only shown for the EPS09
 group \cite{Eskola:2009uj}.}
\end{figure}
%
In Fig.\ \ref{fig:3}, we plot the total parton densities defined in
Eqs.\ (\ref{eq:1}) and (\ref{eq:2}) for protons bound in lead ions
and as obtained in the EPS09 (full) \cite{Eskola:2009uj}, HKN07
(dashed) \cite{Hirai:2007sx}, DSSZ (dotted) \cite{deFlorian:2011fp} and
nCTEQ (dot-dashed) \cite{Schienbein:2009kk} fits. As it is usually done,
we plot $x$ times the
nPDF, i.e.\ the momentum distribution of the partons in the bound proton.
Error bands are only shown for the EPS09 analysis. In descending order of
importance,
the gluon, up- and strange-quark densities are shown separately. Those
for down quarks differ only very little from the up-quark densities,
mostly in the valence (large-$x$) region, and are not shown separately.
Whereas the up- (and down-) quark densities show the characteristic
valence behavior and dominate at large $x$, the gluon density becomes
already dominant for $x$-values below $0.2$ and in particular by more
than an order of magnitude in the low-$x$ region around $10^{-3}$
relevant for the prompt-photon data at low $p_T$. The uncertainty estimated
by EPS09 amounts there to almost a factor of two. Our calculations
show that the uncertainty due to the proton PDF alone accounts for less than half
of the uncertainty. If also the other parametrisations are taken into account
in order to eliminate the theoretical bias, the uncertainty increases to more than
a factor of two, due in particular to the different shapes of the HKN07 and nCTEQ gluon fits.
We stress again that we are considering here a region of extrapolation, where
no experimental constraints are taken into account. In the sea-quark
region at low $x$, the up- and strange-quark densities follow roughly
the shape of the gluon as expected, but are smaller by about
an order of magnitude. The uncertainties for the up quark (and down quark)
are somewhat smaller than those of the gluon. This is probably due to better
constraints on the large-$x$ (valence) region, which influence the low-$x$ region
through the sum rules. However, the uncertainty for the strange quark amounts again
to at least a factor of two.
We therefore expect a considerable impact of the nPDF uncertainties,
in particular from the gluon and from the different parametrisations,
on the description of the low-$p_T$ prompt photon data.

\section{Fragmentation function uncertainties}
\label{sec:4}

%
\begin{figure}
 \centering
 \epsfig{file=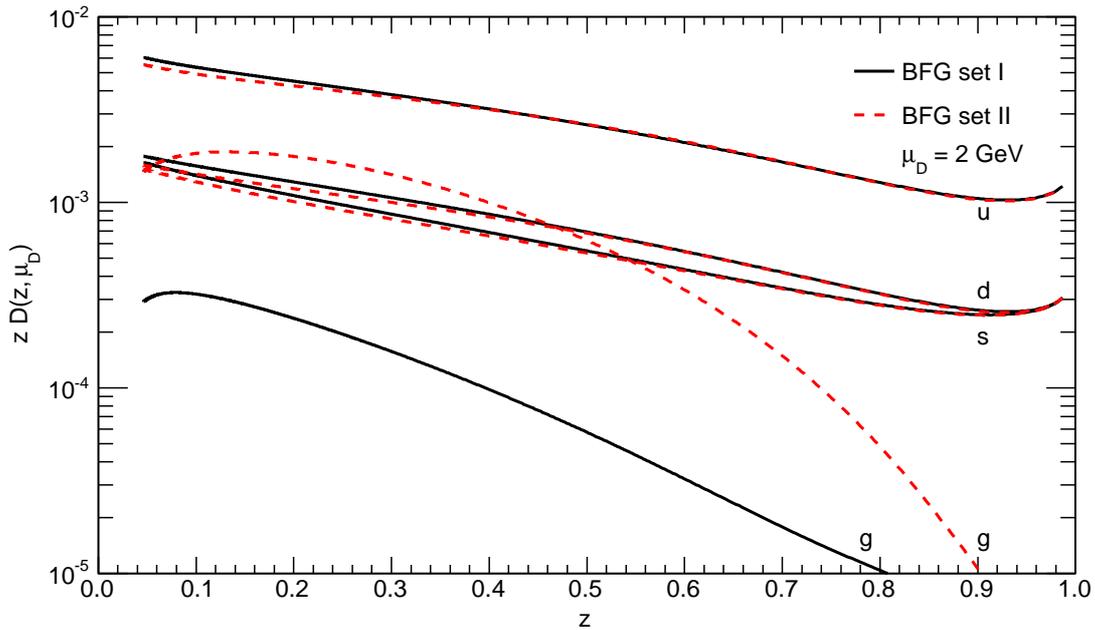,width=\columnwidth}
 \caption{\label{fig:4}Photon fragmentation functions (FFs) for up, down and
 strange quarks as well as for gluons by the BFG collaboration \cite{Bourhis:1997yu}
 at the factorisation
 scale $\mu_f = 2$ GeV. The two sets differ in particular with respect
 to the gluon.}
\end{figure}
%
Fragmentation photons typically contribute less than 20\% to the total yield
in fixed-target
experiments, but can become dominant at collider energies, in particular
at low transverse momenta and in the absence of isolation criteria, as
shown in Sec.\ \ref{sec:2}. Note that the removal of any isolation
criterion is imperative in the search for thermal photon sources in
heavy-ion collisions.
The hadronic input needed for the non-perturbative fragmentation functions
would best be determined from inclusive photon production in $e^+e^-$ annihilation,
but unfortunately the experimental data are very limited and are furthermore
dominated by the pointlike quark-photon fragmentation function.
Therefore, all existing parametrisations use vector-meson dominance to model the
photon fragmentation at low scales. The two most recent BFG parametrisations
have, e.g., been obtained from ALEPH and HRS data on $\rho$ production
\cite{Bourhis:1997yu}. They leave a large uncertainty in the gluon
fragmentation to photons, which is parametrised with the two sets I and II.
Heavy quarks are included above their production thresholds with zero boundary
conditions.

Fig.\ \ref{fig:4} shows the FFs as obtained in the two BFG sets. Again, we
show the momentum distribution, i.e.\ $z$ times the FF $D(z,\mu_D)$ with
$\mu_D=2$ GeV as the factorisation scale. In descending size, the up-,
down- and strange-quark FFs are shown separately, together with the one for
the gluon. Due to its larger charge ($e_u=+2/3$), which enters quadratically,
the up-quark contribution exceeds the one for down and strange quarks with
charge $e_d=e_s=-1/3$ by about a factor of four. The difference between
down and strange quarks is due to quark mass effects. The gluon is very
weakly constrained (essentially only by the momentum sum rule), and its uncertainty
amounts to almost an order of magnitude.
Only when comparing the ratio of gluon and up-quark FFs into pseudoscalar mesons
to the one into vector mesons and assuming that they should be of the same
order due to the reduction of non-perturbative effects, BFG come to the
conclusion that set II may be preferable. As is customary, we have
followed BFG in this choice for our baseline, but stress again that the
uncertainty on the gluon FF into photons remains large.

Despite the fact that photons are colour-neutral objects, they may also
be affected by final-state effects, similarly, but perhaps not quite as
strongly as mesons. In particular, the fragmentation of partons into
photons is expected to be sensitive to the presence of a strongly
interacting medium, since the final partons produced in the hard collision
undergo multiple scattering and suffer energy loss from multiple-gluon
radiation during the fragmentation process. This is usually modeled by
rescaling the energy in a medium-modified fragmentation function \cite{Arleo:2006xb}
\beq
 z D_{\gamma/d}(z,\mu_D)=\int_0^{k_\perp(1-z)}\d\eps \, P_d(\eps,k_\perp) \,
 z^\ast D_{\gamma/d}(z^\ast,\mu_D),
\eeq
where $z^\ast=z/(1-\eps/k_\perp)$ and where $P_d(\eps,k_\perp)$ denotes the
probability that the leading parton $d$ with energy $k_\perp$ has lost an
energy $\eps$ in the medium. The soft-gluon emission is usually assumed
to follow a Poisson distribution. Alternatively, an improved Double Log
Approximation (DLA) \cite{Albino:2009hu}, the Modified Leading-Log
Approximation (MLLA) \cite{Borghini:2005em}, or parton shower and
hadronisation models in Monte Carlo generators can be employed.

One would expect the modification of the fragmentation function to be
largest for low values of the momentum fraction $z$, where its hadronic
component is most important, just as the photon structure function is
dominated by its hadronic component at low $x$ \cite{Albino:2002ck}.
As discussed above, the values of $z$ probed
in the ALICE experiment range from $z\sim0.01$ ($0.1$) to 1 at $p_T=2$ (16)
GeV, i.e.\ we expect the largest modification at low values of $p_T$.
This is also confirmed by a simple argument about the fragmentation time,
which can be estimated with $\hbar/\mu_D\geq0.1$ fm$/c$ for $p_T\leq2$ GeV
and thus becomes comparable to the formation time of the QGP $\tau_0 =
0.15$ $...$ 0.6 fm$/c$ \cite{d'Enterria:2005vz} only at low values of $p_T$.

Since nothing is known experimentally about light vector-meson (in particular $\rho$)
fragmentation in the medium (and very little {\em in vacuo}, see above), one
must resort to pseudoscalar (pion) fragmentation. It is then possible
to simulate both initial- (isospin effects, shadowing) and final-state
effects on inclusive photon production and compare, e.g., to PHENIX
data \cite{Adler:2005ig}. The fragmentation modification is then
found to be strongest at low values of $p_T$ (down to 4 GeV), but the
experimental errors are still too large to identify it unambiguously.
It has therefore been proposed to study not only inclusive photon
production, but also photon-pion correlations to improve the separation
of initial- and final-state effects  \cite{Arleo:2006xb}.

\section{Direct-photon production in heavy-ion collisions}
\label{sec:5}

In this section, we compute the transverse-momentum distribution
of prompt (non-thermal) photons in NLO QCD and compare them to the
experimental data for direct photons (not coming from decays) of the
ALICE collaboration. Particular emphasis is put on the theoretical
errors coming from the different sources described above, i.e.\ an
individual variation of the renormalisation and factorisation scales
by a factor of two up and down the central value $p_T$, the nuclear
PDFs from the EPS09 fits as described in Sec.\ \ref{sec:3}, and the
FFs from the BFG I and II fits as described in Sec.\ \ref{sec:4}.
Our goal is to establish to what extent the
excess of the direct-photon data over the perturbative QCD prediction
survives, once it is subtracted and all theoretical errors are taken
into account, and how this procedure and these errors affect the
determination of the effective temperature of the QGP.

%
\begin{figure}
 \centering
 \epsfig{file=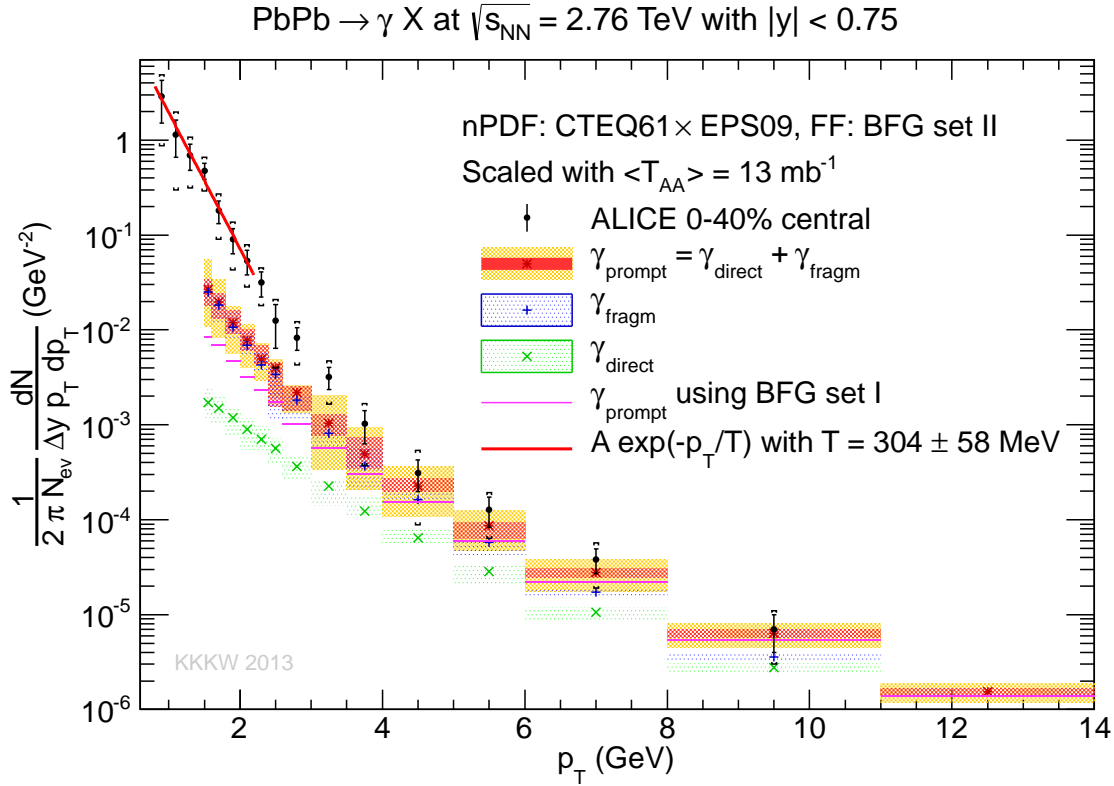,width=\columnwidth}
 \caption{\label{fig:5}Transverse-momentum distribution of direct photons
 in central (0-40\%) lead-lead collisions at $\sqrt{s_{NN}}=2.76$ TeV
 (black data points \cite{Wilde:2012wc}) compared to our NLO QCD predictions
 for prompt photons (red) as well as their fragmentation (blue) and direct
 (green) contributions. Independent scale (yellow) and nPDF (red/blue/green)
 uncertainties as well as the FF variation (magenta) are shown separately, as 
 are the statistical (vertical bars) and systematic (horizontal brackets)
 experimental errors.}
\end{figure}
%
In Fig.\ \ref{fig:5} we compare the transverse-momentum distribution of
the ALICE data obtained in central (0-40\%) lead-lead collisions at
$\sqrt{s_{NN}}=2.76$ TeV (black data points \cite{Wilde:2012wc}) to our
NLO QCD predictions for prompt photons (red) as well as their
fragmentation (blue) and direct (green) contributions.
The centrality of the collision enters our calculations as an overall
normalisation factor \cite{d'Enterria:2003qs}
\beq
 \langle T_{\rm PbPb}\rangle_{0-40\%}\equiv
 {\int_{b_{\min}}^{b_{\max}}\d^2b\,T_{\rm PbPb}
 \over
 \int_{b_{\min}}^{b_{\max}}\d^2b},
 \label{eq:5.1}
\eeq
which relates the invariant differential cross section $E\,\d^3\sigma/\d p^3=
\d\sigma/(2\pi\,\Delta y\,p_T\d p_T)$ (after integration over the azimuthal
angle $\int\d\phi=2\pi$ and for a finite rapidity interval $|y|<0.75$) to
the invariant yield through \cite{d'Enterria:2003qs}
\beq
 {1\over 2\pi \,N_{\rm ev}}{\d N\over\Delta y \,p_T \d p_T} =
 \langle T_{\rm PbPb}\rangle_{0-40\%}
 {\d \sigma\over2\pi \,\Delta y \,p_T \d p_T}.
 \label{eq:5.2}
\eeq
If we use for Eq.\ (\ref{eq:5.1}) the integration boundaries and averaging
integrals of the ALICE experiment given in Tab.\ 1 of Ref.\ \cite{Abelev:2013qoq},
we find $\langle T_{\rm PbPb}\rangle_{0-40\%}=13$ mb$^{-1}$.
If we then apply this normalisation to our theoretical predictions
for the transverse-momentum spectrum in Fig.\ \ref{fig:5} (color) and compare
to the experimental data (black) in the highest measured $p_T$ bin
(8 GeV $<p_T <$ 11 GeV), we find very good agreement
as expected in this perturbative region.

It is well known that the fragmentation contribution (blue) is
considerably softer in $p_T$ than the direct contribution (green),
i.e.\ the former falls off much more steeply than the latter.
In Fig.\ \ref{fig:5}, the fragmentation contribution dominates
in all significant
$p_T$-bins, and in the lowest $p_T$-bins it is more than an order
of magnitude larger than the direct contribution. This leads to a
large uncertainty from the fragmentation function.
E.g., in the lowest $p_T$-bins the theoretical prediction with BFG I FFs
(magenta) is about a factor of three smaller than the one with BFG II FFs
(blue).
On the basis of BFG I FFs, one would thus attribute a larger
excess to thermal photon production than with BFG II FFs.
In the higher $p_T$-bins, the FF uncertainty is smaller due to
the larger contributions of both direct and quark-fragmentation
photons (see Sec.\ \ref{sec:4}).

The scale variation uncertainty (yellow) also turns out to be
important. When all three scales are varied individually, the
cross section changes by more than a factor of two up and down,
much more than if all scales would be varied
simultaneously. For low scales, where the cross section is larger,
one would thus attribute a smaller excess to thermal photon
production. Note that for $p_T<3$ GeV, we vary the factorisation
scales only from $p_T$ to $2p_T$ to stay above the nPDF and FF
starting scales, which means that the true scale uncertainty there is
even larger than the one shown.
Only at larger values of $p_T$, the scale uncertainty becomes
eventually smaller. It could in principle be reduced by
resumming the logarithmic parts of the higher-order corrections
to all orders.
This has been attempted, but unfortunately the corresponding
codes for transverse-momentum \cite{Fink:2001bt} and joint
\cite{Laenen:2000de,Sterman:2004yk} resummation are not publicly
available. The low-$p_T$ region is furthermore known to be
sensitive to the Gaussian smearing parameters entering
the non-perturbative factor required there
\cite{Landry:2002ix}.

The uncertainties arising from the nPDFs (red, green and blue bands)
are less important than the FF and scale errors, but are still
well visible in Fig.\ \ref{fig:5}. Their relative size remains
constant over many low-$p_T$ bins, corresponding to low
momentum fractions $x$ of the partons in the lead nucleus
and the shadowing and free gluon density uncertainties present
there,
but eventually also becomes somewhat smaller at larger $p_T$ or
$x$. This was expected from Sec.\ \ref{sec:3}, since the
shadowing and gluon uncertainties are smaller and the quark
contribution is larger at higher values of $x$. As the nPDF fits
by the other collaborations mentioned in Sec.\ \ref{sec:3} mostly
fall wihin the EPS09 uncertainty band, we do not show the corresponding
predictions here.

Considering statistical (vertical bars) and
systematic (horizontal brackets) experimental as well as all theoretical
errors described above {\em individually}, we find a photon excess
(i.e.\ non-overlapping error bars) for $p_T<2.4$ GeV and the
bin 2.6 GeV $<p_T<$ 3.0 GeV, i.e.\ at smaller $p_T$-values
than the ALICE collaboration,
which did not take into account theoretical uncertainties, but used a larger
theoretical normalisation (see Fig.\ 6 of Ref.\ \cite{Wilde:2012wc}).
Fitting this excess in the range 0.8 GeV $<p_T<$ 2.2 GeV
with an exponential form $A\,\exp(-p_T/T)$, we find an inverse
slope parameter of $T=304$ $\pm$ 58 MeV
(304 $\pm$ 51 GeV in Ref.\ \cite{Wilde:2012wc}), where
our (slightly larger) error has been obtained by adding
statistical and systematic errors in quadrature and where both
turn out to contribute with equal weight. Note, however,
that these results have been obtained from data which include
still both thermal and prompt photon contributions.

%
\begin{figure}
 \centering
 \epsfig{file=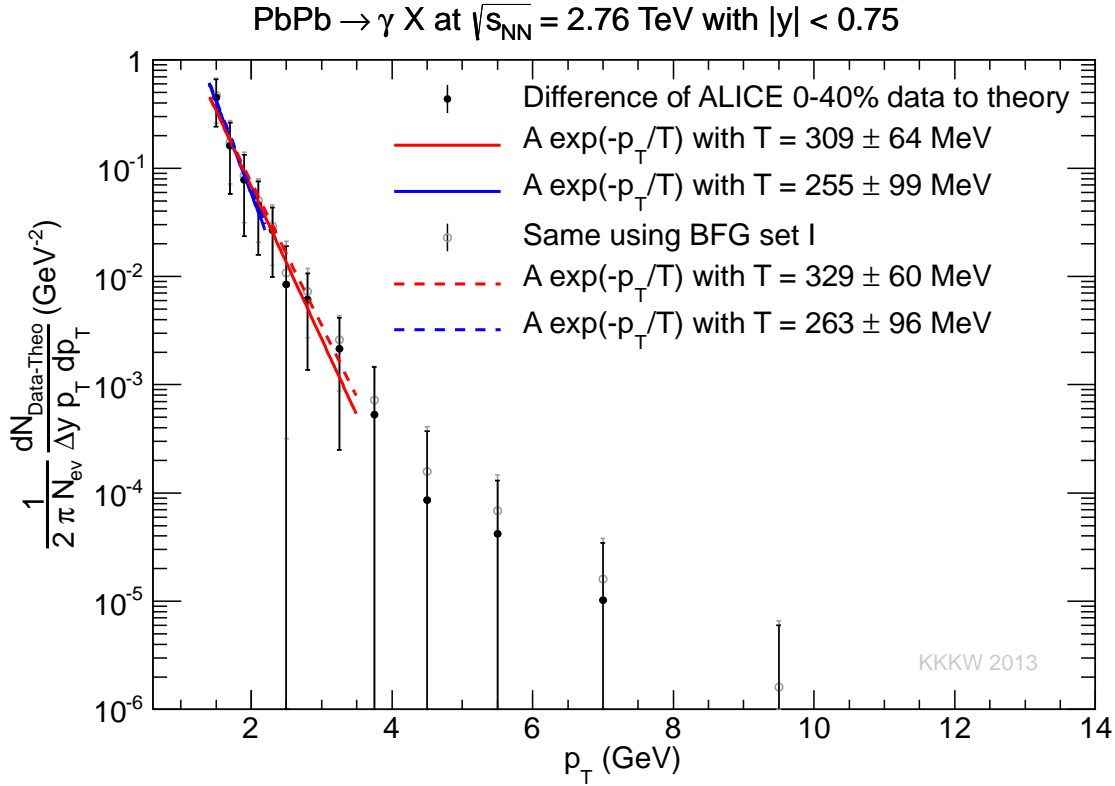,width=\columnwidth}
 \caption{\label{fig:6}Difference of experimental direct-photon and
 theoretical prompt-photon yields using BFG II (full circles)
 and BFG I (open circles) FFs with all other errors added in quadrature.
 Also shown are exponential fits to the points up to $p_T<2.2$ GeV (blue)
 and 3.5 GeV (red).}
\end{figure}
%
The thermal (plus medium-induced) photon production rate alone can be estimated by first
subtracting from the experimental data the theoretically computed
prompt photon contribution and then fitting again the exponential slope
of the difference. This can, however, only be done for $p_T$ values larger
than the starting scales of the nPDFs and FFs. In Fig.\ \ref{fig:6}, the data
points (full circles) now show the difference of data versus
theory. The error bars have been obtained by propagating statistical
and systematic experimental errors as well as scale and nPDF
errors in quadrature. This leads to an additional photon excess
in the bin 3.0 GeV $<p_T<$ 3.5 GeV. The (discrete) FF uncertainty
is estimated by performing the same procedure with the (less
favoured) BFG I fit (open circles) instead of the BFG II fit.
Here, only the experimental errors were taken into account.
Nevertheless, it falls within the total error of the other
uncertainties.

Since the limitation of the previous fit to $p_T<2.2$ GeV is
somewhat arbitrary, we perform two fits to the differential 
data points in Fig.\ \ref{fig:6}. The first one for 1.5 GeV $<p_T<$
2.2 GeV (blue) includes only data points from the overlap of
the experimentally fitted and our theoretically calculated
$p_T$-bins. As the data point in
the lowest $p_T$-bin is somewhat higher than the three data
points below and above it (see Fig.\ \ref{fig:5}),
this fit leads to a lower effective temperature of $255\pm99$ MeV
($263\pm96$ MeV for BFG I FFs). Within its larger error,
it is, however, still in agreement with the unsubtracted
fit value of $304\pm58$ MeV obtained above. If we include all
data points that show a non-zero difference from prompt-photon
production, i.e.\ 1.5 GeV $<p_T<3.5$ GeV, we find instead $309\pm64$ MeV
($329\pm60$ MeV for BFG I FFs),
which is again much closer to the unsubtracted fit value. The additional
data points thus counterbalance the weight of the first one
and bring the result into good agreement with the one based
on Fig.\ \ref{fig:5}.

%
\begin{figure}
 \centering
 \epsfig{file=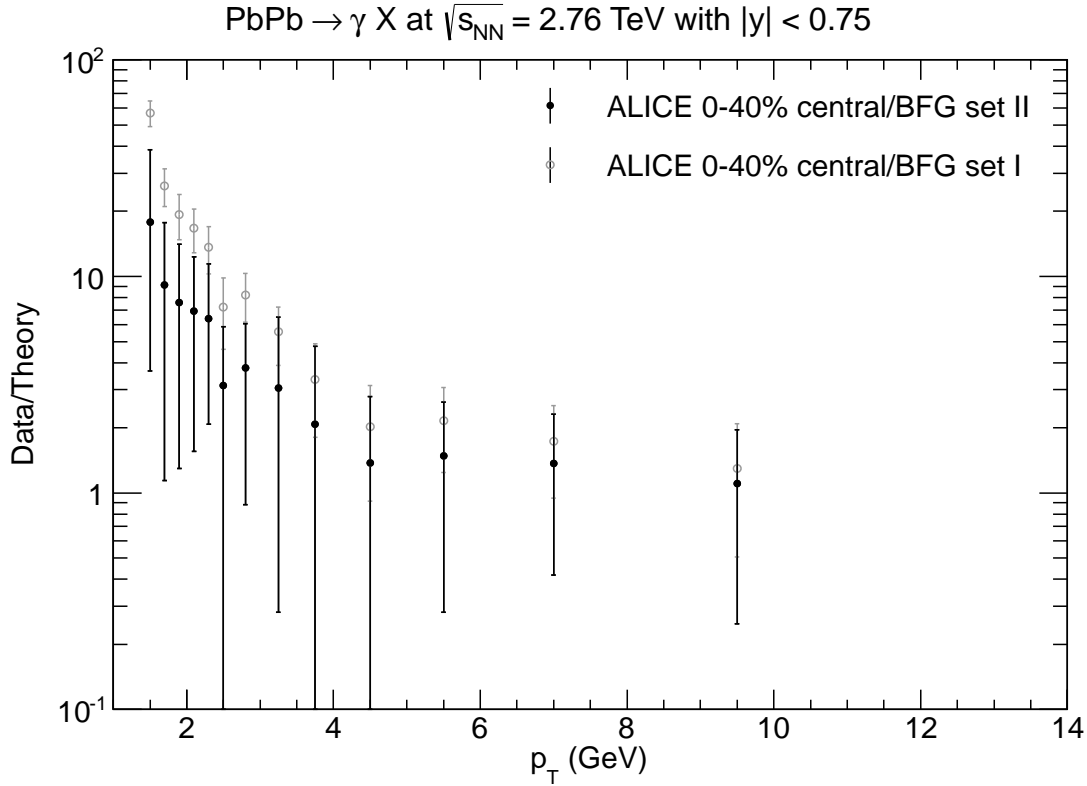,width=\columnwidth}
 \caption{\label{fig:7}Ratio of experimental direct-photon and
 theoretical prompt-photon yields using BFG II (full circles)
 and BFG I (open circles) FFs with all other errors added in quadrature.}
\end{figure}
%
This result may at first sight seem surprising and coincidental.
One must, however, take into account the relative size of the
prompt-photon contribution, which falls off like $p_T^{-4}$
as predicted by our calculations, to
the measured direct-photon rate in the different $p_T$ regions,
shown in Fig.\ \ref{fig:7}. Indeed, we find that prompt photons
contribute less than 20\% for $p_T<2.4$ GeV (less than 10\% for
BFG I FFs), i.e.\ their subtraction and the theoretical error
do not strongly modify the exponential fall-off of the
experimental data in this region. Above 4 GeV,
we find instead an almost constant ratio of prompt over direct
photons that is consistent with one within the uncertainties,
i.e.\ these observed photons are produced in hard scatterings.
In the intermediate region, it was already clear from Fig.\
\ref{fig:5} that the prompt-photon contribution is substantial,
larger than 20\% (10\% for BFG I FFs), and must indeed first be
subtracted from the experimental data. Only then can the excess
be described again by an exponential fit, which we found to be
in good agreement with the fit to the unsubtracted data at low
$p_T$.

Several attempts have been made to describe the transition
from thermal to hard photons in the intermediate $p_T$ region
theoretically. E.g., bremsstrahlung from a fast charged parton may be
induced by the medium \cite{Zakharov:2004bi}, or a hard quark
or gluon and a thermal parton may convert to a hard photon
through annihilation or QCD Compton scattering \cite{Liu:2008zb}.
As discussed above, we found that current data can well be
described by an exponential fit up to $p_T<3.5$ GeV, once the
prompt-photon contribution is subtracted.
If we include the data points above $p_T$ values of 3.5 GeV,
we obtain a very similar effective temperature of $319\pm66$ MeV,
as these data points no longer carry much statistical weight.
An almost identical effective temperature of $311\pm64$ MeV
is also obtained when we vary all scales simultaneously instead
of individually. This clearly demonstrates that the scale variation error is
strongly correlated among different $p_T$-bins. It affects
mostly the normalisation constant $A$ rather than the inverse
slope parameter $T$.

\section{Conclusions}
\label{sec:6}

In conclusion, we have performed in this work a first rigorous theoretical
analysis of the ALICE measurement of direct-photon production in central
lead-lead collisions with a centre-of-mass energy of
$\sqrt{s_{NN}}=2.76$ TeV at low values of $p_T$ of
0.8 to 14 GeV. Based on a next-to-leading order QCD calculation,
we found that prompt photon production in this region was
largely induced by initial gluons and dominated by fragmentation
contributions (for $p_T<4$ GeV those of a final gluon) due to the absence
of an isolation criterion. This resulted in large theoretical uncertainties
from independent variations of the renormalisation and factorisation
scales, nuclear parton densities and fragmentation functions.

Nevertheless, we were able to confirm that the experimental data are well fitted
in the region $p_T\in[0.8;2.2]$ GeV by an exponential form $A\,\exp(-p_T/T)$
with an effective temperature of $T=304$ $\pm$ 58 MeV, with a slightly
larger uncertainty than the 51 MeV quoted by the ALICE collaboration.
The reason is that in this region  prompt photons contribute
less than 10 to 20\% to the total direct photon rate,
so that their subtraction and theoretical error
do not influence the fit result very much. We also verified that already for
values of $p_T>4$ GeV the experimental data fall off with $p_T^{-4}$ as
predicted by perturbative QCD. In this region, the theoretical uncertainty is under
reasonable control and strongly correlated among the different $p_T$-bins,
in particular the one coming from the scale variation.

In the intermediate $p_T$-region from 1.5 to 3.5 GeV, the prompt-photon
contribution had to be subtracted from the experimental data before
a sensible exponential fit could be performed. We were able to verify
an exponential fall-off with a very similar effective temperature
of $309\pm64$ MeV. This result did not change significantly when extended
to $p_T$ values of 7 GeV or when all scales were varied simultaneously,
confirming the strong correlation of the theoretical error among different
$p_T$ bins. Based on current ALICE data, a global thermal description
thus seems to hold also in the intermediate $p_T$ region, and the presence
of additional mechanisms, such as medium-induced bremsstrahlung and/or
thermal parton conversion, do not need to be invoked to reproduce the
measured photon spectra.

\appendix
\section{Prompt-photon yields and their uncertainties in NLO QCD}
\label{sec:a}

For future use, we present in Tab.\ \ref{tab:1} our NLO QCD calculations for the
invariant prompt-photon yield in central (0-40\%) lead-lead
collisions at $\sqrt{s_{NN}}=2.76$ TeV and their uncertainties
as a function of transverse momentum in tabular form. All numbers
have been rescaled according to Eq.\ (\ref{eq:5.2}) using
$\langle T_{\rm PbPb}\rangle_{0-40\%}=13$ mb$^{-1}$.

\begin{table}[tbp]
\centering
\begin{tabular}{c|ccccc|cc|c}
$p_T$ (GeV) & BFG set II & \multicolumn{2}{c}{PDF error} & \multicolumn{2}{c|}{Scale error} & Fragm. & Direct & BFG set I \\
 & & + & - & + & - & & & \\
\hline
1.5 - 1.6 & 2.68e-02 & 7.2e-03 & 8.4e-03 & 2.8e-02 & 1.6e-02 & 2.51e-02 & 1.71e-03 & 8.39e-03 \\
1.6 - 1.8 & 1.98e-02 & 4.7e-03 & 6.4e-03 & 1.4e-02 & 1.1e-02 & 1.83e-02 & 1.51e-03 & 6.90e-03 \\
1.8 - 2.0 & 1.19e-02 & 4.3e-03 & 2.8e-03 & 5.9e-03 & 6.2e-03 & 1.07e-02 & 1.19e-03 & 4.66e-03 \\
2.0 - 2.2 & 7.77e-03 & 2.4e-03 & 1.7e-03 & 3.7e-03 & 3.8e-03 & 6.87e-03 & 8.99e-04 & 3.22e-03 \\
2.2 - 2.4 & 4.96e-03 & 1.9e-03 & 6.8e-04 & 2.3e-03 & 2.0e-03 & 4.26e-03 & 7.05e-04 & 2.32e-03 \\
2.4 - 2.6 & 3.98e-03 & 5.0e-04 & 2.4e-03 & 8.7e-04 & 2.0e-03 & 3.41e-03 & 5.68e-04 & 1.72e-03 \\
2.6 - 3.0 & 2.20e-03 & 3.6e-04 & 7.8e-04 & 4.4e-04 & 8.5e-04 & 1.83e-03 & 3.66e-04 & 1.01e-03 \\
3.0 - 3.5 & 1.04e-03 & 2.5e-04 & 2.7e-04 & 1.0e-03 & 7.0e-04 & 8.15e-04 & 2.27e-04 & 5.73e-04 \\
3.5 - 4.0 & 4.94e-04 & 2.4e-04 & 1.5e-04 & 4.4e-04 & 2.9e-04 & 3.70e-04 & 1.23e-04 & 3.06e-04 \\
4.0 - 4.5 & 2.87e-04 & 7.4e-05 & 3.6e-05 & 1.8e-04 & 1.5e-04 & 2.09e-04 & 7.80e-05 & 1.91e-04 \\
4.5 - 5.0 & 1.72e-04 & 3.0e-05 & 2.5e-05 & 9.6e-05 & 9.0e-05 & 1.19e-04 & 5.22e-05 & 1.21e-04 \\
5.0 - 5.5 & 1.09e-04 & 9.8e-06 & 4.0e-05 & 4.6e-05 & 5.2e-05 & 7.33e-05 & 3.53e-05 & 7.37e-05 \\
5.5 - 6.0 & 6.58e-05 & 1.2e-05 & 1.5e-05 & 3.5e-05 & 2.8e-05 & 4.35e-05 & 2.23e-05 & 4.64e-05 \\
6.0 - 6.5 & 4.65e-05 & 7.4e-06 & 5.6e-06 & 1.5e-05 & 2.0e-05 & 3.01e-05 & 1.64e-05 & 3.49e-05 \\
6.5 - 7.0 & 3.05e-05 & 6.1e-06 & 3.1e-06 & 1.3e-05 & 1.0e-05 & 1.87e-05 & 1.18e-05 & 2.50e-05 \\
7.0 - 7.5 & 2.28e-05 & 1.4e-06 & 7.7e-06 & 5.8e-06 & 8.6e-06 & 1.42e-05 & 8.55e-06 & 1.71e-05 \\
7.5 - 8.0 & 1.50e-05 & 3.5e-06 & 1.5e-06 & 7.0e-06 & 4.4e-06 & 8.50e-06 & 6.54e-06 & 1.31e-05 \\
8.0 - 8.5 & 1.21e-05 & 9.5e-07 & 2.3e-06 & 3.5e-06 & 3.8e-06 & 7.12e-06 & 4.94e-06 & 9.84e-06 \\
8.5 - 9.0 & 8.74e-06 & 1.0e-06 & 1.5e-06 & 1.9e-06 & 2.7e-06 & 5.16e-06 & 3.59e-06 & 7.31e-06 \\
9.0 - 9.5 & 6.47e-06 & 1.6e-06 & 9.8e-07 & 2.1e-06 & 1.8e-06 & 3.51e-06 & 2.97e-06 & 5.78e-06 \\
9.5 - 10.0 & 5.00e-06 & 1.6e-06 & 3.8e-07 & 1.3e-06 & 1.4e-06 & 2.68e-06 & 2.31e-06 & 4.42e-06 \\
10.0 - 10.5 & 4.10e-06 & 4.4e-07 & 7.0e-07 & 1.0e-06 & 1.1e-06 & 2.16e-06 & 1.94e-06 & 3.58e-06 \\
10.5 - 11.0 & 3.13e-06 & 6.0e-07 & 2.9e-07 & 7.8e-07 & 7.1e-07 & 1.75e-06 & 1.38e-06 & 2.71e-06 \\
11.0 - 11.5 & 2.48e-06 & 3.0e-07 & 3.1e-07 & 5.1e-07 & 6.5e-07 & 1.30e-06 & 1.17e-06 & 2.28e-06 \\
11.5 - 12.0 & 2.00e-06 & 2.7e-07 & 3.3e-07 & 4.9e-07 & 5.0e-07 & 1.08e-06 & 9.21e-07 & 1.76e-06 \\
12.0 - 12.5 & 1.63e-06 & 3.5e-07 & 3.1e-08 & 4.2e-07 & 3.5e-07 & 8.40e-07 & 7.92e-07 & 1.45e-06 \\
12.5 - 13.0 & 1.36e-06 & 7.9e-08 & 2.4e-07 & 2.5e-07 & 3.1e-07 & 6.96e-07 & 6.61e-07 & 1.23e-06 \\
13.0 - 13.5 & 1.13e-06 & 3.5e-08 & 2.9e-07 & 2.0e-07 & 3.0e-07 & 5.72e-07 & 5.53e-07 & 1.03e-06 \\
13.5 - 14.0 & 8.89e-07 & 2.2e-07 & 1.3e-07 & 2.4e-07 & 1.8e-07 & 4.57e-07 & 4.32e-07 & 8.29e-07 \\
14.0 - 14.5 & 7.78e-07 & 9.2e-08 & 8.5e-08 & 1.8e-07 & 1.7e-07 & 3.82e-07 & 3.96e-07 & 7.35e-07 \\
14.5 - 15.0 & 6.51e-07 & 7.5e-08 & 6.6e-08 & 1.1e-07 & 1.5e-07 & 3.17e-07 & 3.34e-07 & 6.00e-07 \\
15.0 - 15.5 & 5.09e-07 & 1.2e-07 & 2.8e-08 & 1.1e-07 & 8.0e-08 & 2.44e-07 & 2.65e-07 & 4.84e-07 \\
15.5 - 16.0 & 4.69e-07 & 3.0e-08 & 1.3e-07 & 5.5e-08 & 1.1e-07 & 2.41e-07 & 2.28e-07 & 4.13e-07 \\
\end{tabular}
\caption{\label{tab:1}Invariant prompt-photon yields and their uncertainties in NLO QCD.}
\end{table}

\acknowledgments

The work of C.\ Klein-B\"osing was supported by the Helmholtz Alliance
Program of the Helmholtz Association, contract HA216/EMMI ``Extremes of
Density and Temperature: Cosmic Matter in the Laboratory''.
We thank the ALICE collaboration for making their preliminary
data available to us and J.P.\ Guillet, J.\ Owens, W.\ Vogelsang
and M.\ Wilde for useful discussions.




\end{document}